\documentclass[preprint,12pt]{elsarticle}

\journal{Cryogenics}

\begin{document}

\begin{frontmatter}



\title{High Performance Heat Conductor with Small Spring Constant for Cryogenic Applications}


\author[label1]{Tomohiro Yamada}
\address[label1]{Research Center for Cosmic Neutrinos, Institute for Cosmic Ray Research (ICRR), The University of Tokyo, 5-1-5, Kashiwanoha, Kashiwa, Chiba, 277-8582, Japan}

\author[label2,label3]{Takayuki Tomaru}
\address[label2]{Kamioka Branch, National Astronomical Observatory of Japan (NAOJ), 238, Higashimozumi, Kamioka-cho, Hida City, Gifu 506-1205, Japan}
\address[label3]{KAGRA Observatory, Institute for Cosmic Ray Research (ICRR), The University of Tokyo, 5-1-5, Kashiwanoha, Kashiwa, Chiba, 277-8582, Japan}

\author[label3]{Toshikazu Suzuki}

\author[label4]{Takafumi Ushiba}
\address[label4]{KAGRA Observatory, Institute for Cosmic Ray Research (ICRR), The University of Tokyo, 238, Higashimozumi, Kamioka-cho, Hida City, Gifu 506-1205, Japan}

\author[label3,label5]{Nobuhiro Kimura}
\address[label5]{Cryogenics Science Center, Applied Research Laboratory, High Energy Accelerator Research Organization, 1-1, Oho, Tsukuba, Ibaraki, 305-0801, Japan}

\author[label6]{Suguru Takada}
\address[label6]{Device Engineering and Advanced Physics Research Division, National Institute for Fusion Science, 322-6, Oroshi, Toki, Gifu, 509-5292, Japan}

\author[label7]{Yuki Inoue}
\address[label7]{Department of Physics, National Central University, No. 300, Zhongda Rd., Zhongli District, Taoyuan, 32001, Taiwan}

\author[label1]{Takaaki Kajita}

\begin{abstract}

We developed a soft and high thermal conductive heat link for cryogenic applications. The measured maximum thermal conductivity was approximately 18500$\, \rm W/m/K$ at 10$\,$K.  This spring constant was 1/43 of that of a single thick wire with the same cross-sectional area at room temperature. We realized these performances by utilizing high purity aluminum (99.9999$\,\%$, 6$\,$N) and by stranding many thin wires that are 0.15$\,$mm in diameter. The electrical Residual Resistivity Ratio (RRR) was also measured and used to estimate the thermal conductivity by Wiedemann-Franz law. We observed the size effect on the electrical conductivity, and evaluated the effect of mechanical deformation to the RRR. These evaluations allow us to design a pragmatic heat link for cryogenic application.
\end{abstract}


\end{frontmatter}


\section{Introduction}
\label{Introduction}

Cryogenic environments are useful for high precision measurements as they reduce the effects of thermal fluctuation. It is quite remarkable that nowadays mechanical cryocoolers have become increasingly popular and by using them, we can easily reach to liquid helium temperatures. However, since cryocoolers have mechanically moving parts to exchange working gas, the vibration of the cold head sometimes becomes a problem in the vibration sensitive experiments. This problem had actually been observed in the KAGRA experiment in Japan [1-3], hence, we needed to develop a soft and high thermal conductive heat conductor that could be put to use even at cryogenic temperatures.

KAGRA is a laser interferometer based gravitational wave detector constructed in the Kamioka mine of Japan. One key feature of KAGRA is that it has cryogenic mirror systems to reduce thermal noise of mirrors and suspensions caused by thermal fluctuation [4]. We need to cool down the 200$\,$kg mirror suspension system to 20$\,$K in an ultra-high vacuum condition. Since radiative cooling works effectively only in high temperature regions, conductive cooling must be used in the low temperature region, or in temperatures typically below the liquid nitrogen temperature. At the same time, we need to pay attention to the vibration from the cryocooler through the heat conductor, because KAGRA is extremely sensitive to mechanical vibrations.

A previous study showed an idea to solve this problem, where many thin wires consisting of high purity metal were used as a thermal conductor [5,6]. High purity metals like aluminum and copper have very high thermal conductivities at cryogenic temperatures. In particular, aluminum has almost half the value of the Young's modulus for copper, and for this reason, we can expect to realize a soft heat link by adopting this material. Therefore, we developed a pure aluminum heat link for this purpose.

\section{Theory}
\subsection{Theory of Thermal Conductance of Metals at Low Temperature}
\label{Theory of thermal conductance of metals at low temperature}

Electrons are the primary heat carriers in a metal [7]. The thermal conductivity $\kappa$ of the metal is generally written as
\begin{equation}
\kappa = \frac{1}{3}c_{\scalebox{0.5}{V}}vl,
\end{equation}
\noindent where $c_{\scalebox{0.5}{V}}$ is the specific heat at constant volume, and $v$ and $l$ are the velocity and the mean-free path of conduction electron, respectively. In very low temperature regions, the thermal conductivity is proportional to $T$ for the following three reasons: 1) the specific heat is proportional to temperature $T$, 2) the velocity of electron is almost constant since energy of electrons is determined by Fermi energy, 3) the mean-free path of electrons is constant since it becomes comparable size with density of impurities and defects in metal at low temperatures. Thus, high purity metals have larger conductivities at low temperatures. In the case of a high purity metal with a small diameter, the diameter of a specimen can be comparable in size with a mean-free path of an electron and can limit the conductivity. This is known as the size effect. In very low temperature regions, the Wiedemann-Franz law holds;

\begin{equation}
\label{Wiedemann-Franz law}
\frac{\kappa}{\sigma} = \frac{\pi^{2}}{3}  \left(\frac{k_{\rm B}}{e} \right)^{2} T,
\end{equation}

\noindent where $\kappa$ and $\sigma$ are the thermal and electrical conductivities, respectively, $k_{\rm B}$ is the Stefan-Boltzmann constant and $e$ is the elementary charge. A coefficient of $T$ is called the Lorenz factor, and theoretically, does not depend on the type of material, while practically, slightly depends on the material [8]. This law tells us that there is a relationship between the thermal and electrical conductivities.

Thermal conductivity measurement generally requires a long time. On the other hand, electrical conductivity measurement is much easier, and as a convenient alternative to actually performing thermal conductivity measurement, we can estimate thermal conductivity from the result of the electrical conductivity measurement and eq.(\ref{Wiedemann-Franz law}). Here we demonstrate that the Residual Resistivity Ratio (RRR) is a good parameter to evaluate electrical conductivity, which is defined as,

\begin{equation}
{\rm RRR} = \frac{\rho_{300\rm K}}{\rho_{4.2\rm K}} = \frac{\sigma_{4.2\rm K}}{\sigma_{300\rm K}} ,
\label{RRR_definition}
\end{equation}

\noindent where $\rho$ is the electrical resistivity. To estimate the thermal conductivity at higher temperature regions than that of which the Wiedemann-Franz law holds, we need to consider the electron scattering by phonons. Since the number of phonons is proportional to $T^{3}$, the mean-free path of conduction electrons has the temperature dependence of $T^{-3}$. Since the specific heat is proportional to $T$, the thermal conductivity $\kappa$ is proportional to $T^{-2}$. By combining the temperature dependence of thermal conductivity at two temperature ranges, we can express the thermal conductivity as

\begin{equation}
\kappa = \frac{1}{AT^{2} + B/T}
\label{RRR_tc_estimation}
\end{equation}

\noindent at low temperature. The first term represents the scattering of conduction electrons by phonons, and the second term comes from the Wiedemann-Franz law. Coefficients A and B depend on the type of material. Previous studies have reported as A$\,=1.8 \times 10^{-7} $ and B$\,= 1.1/\rm{RRR}$ for pure aluminum. This expression can be applied below $\theta /10$ ($\theta$ is the Debye temperature) and this is approximately 30$\,$K in the case of aluminum.

\subsection{Theory of Stiffness}
\label{Theory of stiffness}

The Young's modulus is an important parameter for evaluating the vibration transfer in the link. The spring constant of a cylindrical wire, $k$, is generally written as 

\begin{equation}
k = K \frac{EI}{L^{3}} = K \frac{E}{L^{3}} \, \frac{\pi r^{4}}{4},
\end{equation}

\noindent where $K$ is the shape factor, $I$ is the moment of inertia of area, $L$ is the length of a wire and $r$ is the radius of a wire. In the case of a cantilever shape, $K = 3$. When the number of a wire is $N$, this situation is equal to the case of parallel $N$ springs. The spring constant of parallel $N$ springs $k_{N}$ is written as

\begin{equation}
k_{N} = N \times k_{1} = N \times K \frac{E}{L^{3}} \, \frac{\pi r_{1}^{4}}{4}.
\end{equation}

\noindent Here, $k_{1}$ and $r_{1}$ are the spring constant and the radius of a single wire. When the total cross-sectional area of $N$ wires and a thick single wire are the same, we can derive the relationship between $k_{N}$ and $k$ as

\begin{equation}
k_{N} = \frac{1}{N}\, k.
\label{equation_springconstant_decrease}
\end{equation}

\noindent Therefore, it is obvious that we can effectively decrease the spring constant by using many thin wires. Then, the resonant frequency, $f$,  is proportional to the square root of the spring constant;

\begin{equation}
f_{N} = \frac{1}{\sqrt{N}}\, f.
\label{equation_springconstant_decrease_freq}
\end{equation}

\section{Development of the Heat Link}
\label{Development of the heat link}

In this section, we introduce the details of the developed heat link. We adopted high purity aluminum with 99.9999$\,$\% purity (6$\,$N) for the material to balance thermal conductance and softness. We fabricated a stranded cable consisting of 49 thin single wires with 0.15$\,$mm diameter, where seven single wires were used to form the primary twist and were gathered together to form the seven secondary twist. The pure aluminum was produced by Sumitomo Chemical Co., Ltd., thin wires were made by Noge Electric Industries Co., Ltd. and the stranding work was done by Arai Co., Ltd.

\begin{figure}[h]
\begin{center}
 \includegraphics[width=18pc]{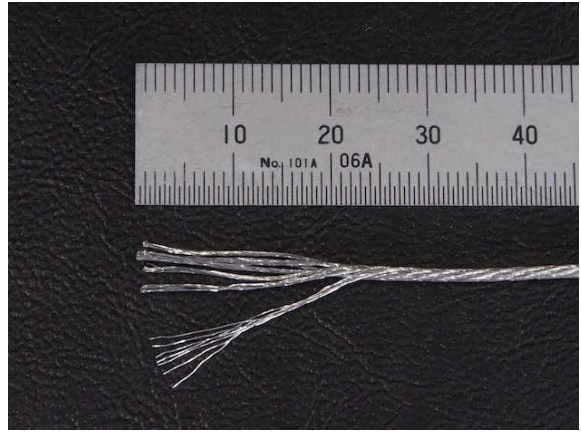}
 \caption{Details of the developed heat link. Outer diameter of strand is approximately 1$\,$mm. Left tip is loosened to show details}
 \label{heatlink_strands.jpg}
\end{center}
\end{figure}

\section{Experiment}
\label{Experiment}

\subsection{RRR}
\label{RRR}

\subsubsection{RRR Measurement}
\label{RRR measurement}

Table$\,$\ref{RRR3} shows a typical RRR data in literatures and estimated mean-free path of an electron. Here we measured the RRRs of 6$\,$N and 5$\,$N specimens. The size of measured specimens were 0.15$\,$mm in diameter and about 300$\,$mm in length. All samples were annealed in vacuum for 3 hours at $450\,^{\circ}$C. We adopted the 4-wire measurement method to measure the electrical resistance precisely.

\newpage
\begin{table}[h]
  \begin{center}
    \begin{tabular}{|c||c|c|c|} \hline
            &$\rm RRR_{bulk}$ &$\rm RRR_{0.5mm}$& $\,\,\, \lambda _{\rm bulk}$ \,\,\, \\ \hline \hline
      6N &  22000                  &   12500                     &   0.7$\,$mm                               \\ \cline{1-4}
      5N &  6000                    &     -                            &   0.2$\,$mm                               \\ \cline{1-4}
    \end{tabular}
    \caption{RRR of bulk [9] and plate with 0.5$\,$mm thickness [6]. $\lambda _{\rm bulk}$ shows estimated mean-free path of conduction electron for bulk.[9]}
    \label{RRR3}
  \end{center}
\end{table}

Table$\,$\ref{RRR4} shows the measured results. The measured RRR for a 6$\,$N wire with 0.15$\,$mm diameter was 6 times and 1.7 times smaller than that of a 6$\,$N bulk and 5$\,$N bulk, respectively. And this value was only 1.2 times larger than that of a 5$\,$N wire. The reduction of RRRs in the case of a thin wire can be explained by size effect, which is because the mean-free path of conduction electron for 6$\,$N bulk is 0.7$\,$mm in estimation and this is much larger than the wire diameter.

\begin{table}[h]
  \begin{center}
    \begin{tabular}{|c||c|c|c|} \hline
            &Elect. Resis.&$\rm RRR_{wire}$& Calculation\\ \hline \hline
      6N & 132$\,\rm\mu\Omega$ &3600&  3882         \\ \cline{1-4}
      5N & 154$\,\rm\mu\Omega$ &3100&  2571         \\ \cline{1-4}
    \end{tabular}
    \caption{Measured electrical resistance of 300$\,$mm specimen with 0.15$\,$mm diameter at 4.2$\,$K, measured RRR of wire and estimated RRR by eq.(\ref{RRR_measurement}).}
    \label{RRR4}
  \end{center}
\end{table}

The RRR of a thin wire is roughly calculated by

\begin{equation}
\label{RRR}
{\rm RRR}^{-1}_{\rm wire} = {\rm RRR}^{-1}_{\rm bulk} + \left ({\rm RRR}^{-1}_{\rm bulk} \cdot \lambda _{\rm bulk} \right ) \cdot d^{-1},
 \label{RRR_measurement}
\end{equation}

\noindent where ${\rm RRR}_{\rm wire}$ and ${\rm RRR}_{\rm bulk}$ are the RRRs of a thin wire and bulk, respectively [10]. A $\lambda _{\rm bulk}$ is the mean-free path of a conduction electron and $d$ is the diameter of a thin wire. The RRRs have been calculated by using eq.(\ref{RRR_measurement}), which resulted in 3882 for 6$\,$N and 2571 for 5$\,$N. These results support our consideration from the theoretical aspect.

\subsubsection{Influence of Mechanical Deformation to the RRR}
\label{Influence of Mechanical Deformation to the RRR}

In practical use, it is difficult to avoid deformation of the heat link. Distortion inside the heat link as a result of deformation causes reduction of electrical and thermal conductivities. In this experiment, we investigated the effect of deformation on the RRR by wrapping a specimen around a cylindrical bar. We wrapped a thin wire with 0.15$\,$mm diameter and 1$\,$g weight around cylindrical bars. This corresponds to a wire receiving a tension of 0.55$\,$$\rm MPa$. Tensile strength of pure aluminum is about 100$\,$$\rm MPa$, therefore, longitudinal stress is negligible. 

We measured the RRR of a thin wire, and after wrapping it ten times around cylindrical bar, we measured its RRR again. We investigated the dependence of internal strain by using cylinderical bars of diffenrent sizes, varying between 10$\,$mm and 40$\,$mm. We measured 5 specimens for each diameter.

\begin{figure}[h]
 \begin{minipage}{0.47\hsize}
  \begin{center}
   \includegraphics[width=14pc]{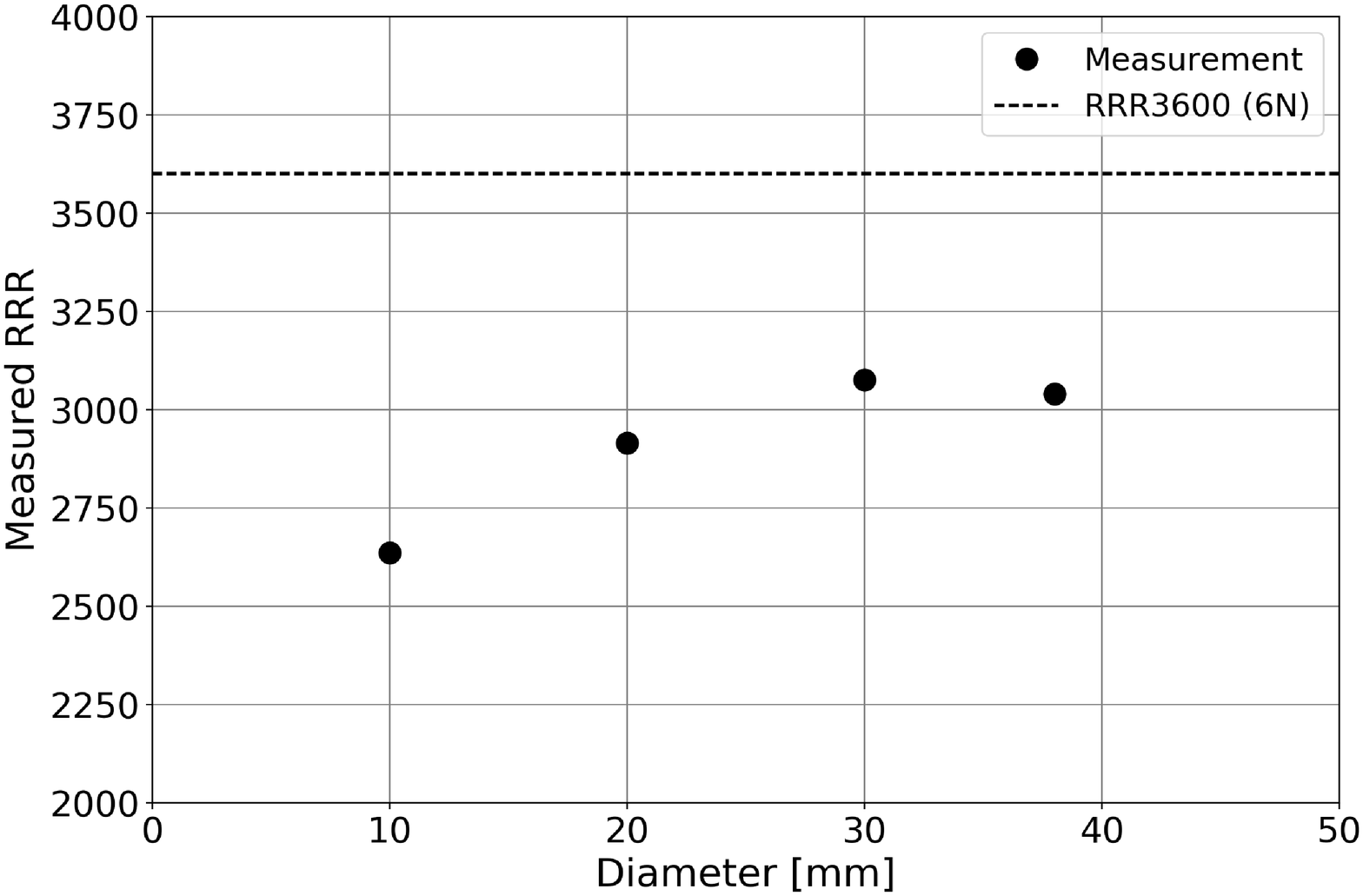}
   \caption{Relationship between the diameter of a cylindrical bar and measured RRR. A dashed line shows the data of a specimen before being wrapped around bars.}
   \label{RRR_deformation.png}
  \end{center}
 \end{minipage}
 \begin{minipage}{0.04\hsize}
  \hspace{2mm}
 \end{minipage}
 \begin{minipage}{0.47\hsize}
  \begin{center}
   \includegraphics[width=14pc]{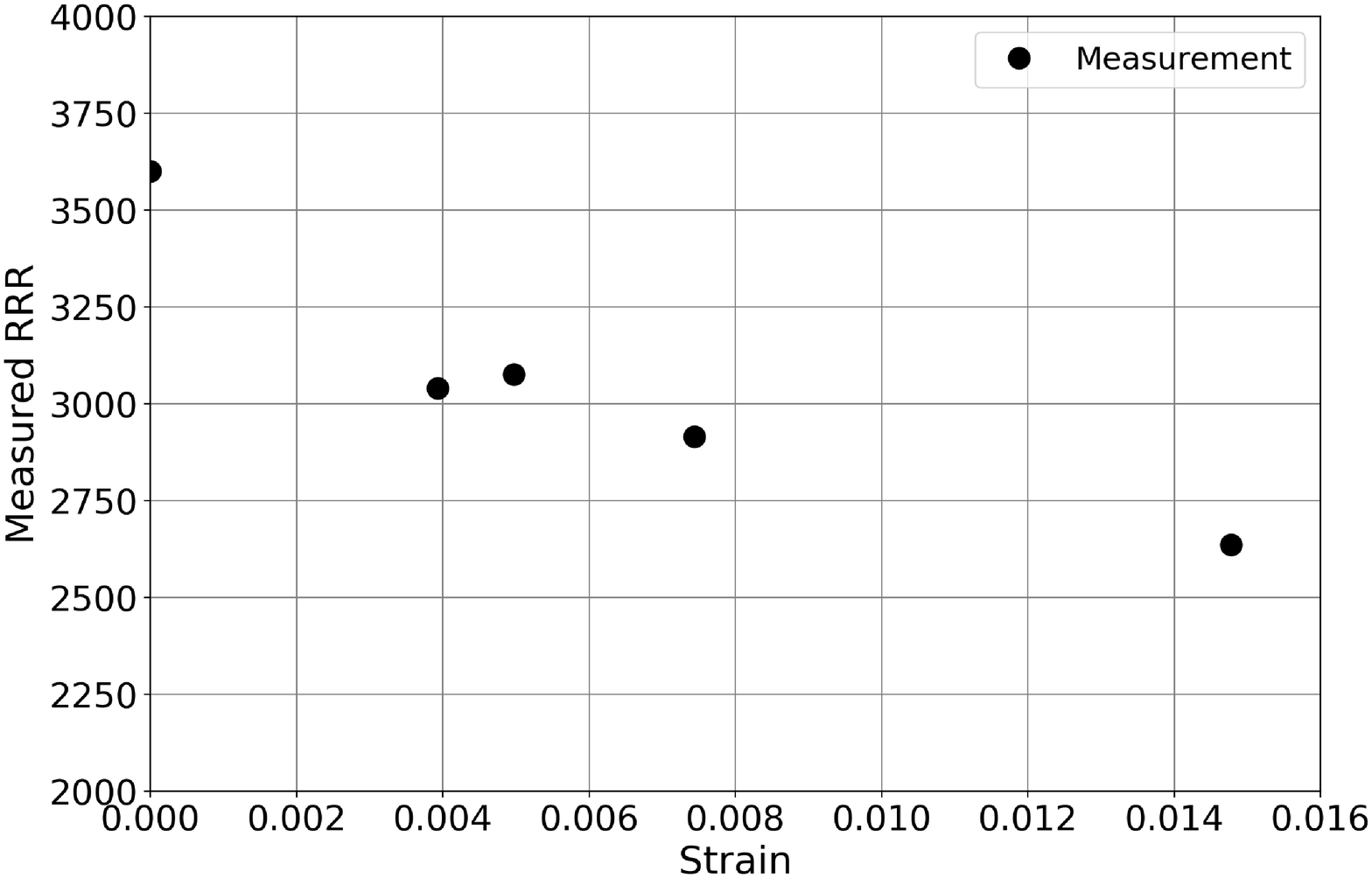}
   \caption{Relationship between strain and measured RRR. Wrapping around smaller diameter of cylinder causes larger strain. Strain of 0 means the data of a specimen before being wrapped around bars.}
   \label{RRR_strain}
  \end{center}
 \end{minipage} 
\end{figure}

Figure$\,$\ref{RRR_deformation.png} shows the relationship among the diameter of a cylindrical bar and measured RRRs. For example. for the case of a 10$\,$mm cylinder, the RRR decreased 30$\,\%$. Figure$\,$\ref{RRR_strain} shows RRR dependence of strain, which is converted from diameter. The RRR became small when the diameter was small.

As we described in Section$\,$\ref{Theory of thermal conductance of metals at low temperature}, there is a relationship between the RRR and the thermal conductivity. So we expect to see a similar dependence of strain for RRR with thermal conductivity. 

\subsection{Thermal Conductivity}
\label{Thermal Conductivity}

Developing a heat link with large thermal conductivity will allow us to reduce the required number of heat links. This means that we can expect to reduce the amount of vibration transmitted to experimental devices through the heat link.

\begin{figure}[h]
 \centering
 \includegraphics[width=15pc]{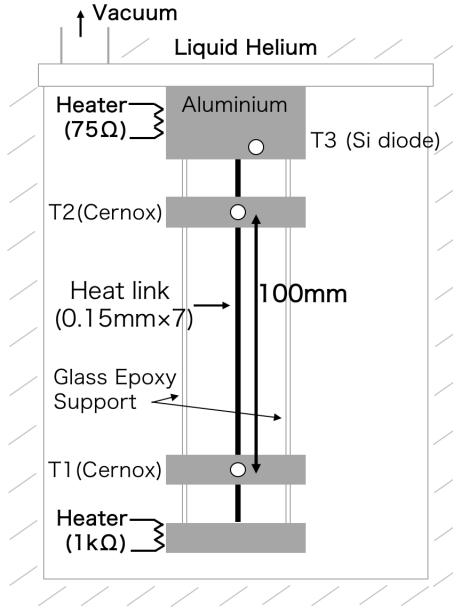}
 \caption{Setup of thermal conductivity measurement. Grey  rectangles represent the aluminum blocks where heaters and thermometers are mounted. Top aluminum block is directly in contact with the lid of vacuum tank. Two glass epoxy fibers support all aluminum blocks so that the specimens are not stressed due to their weight.}
 \label{tc_setting.png}
\end{figure}

We measured the thermal conductivity of a strand of 7 thin wires. We used the longitudinal heat flow method (see Figure$\,$\ref{tc_setting.png}) [11]. The temperature gradient was given by a metal film resister ($\rm 1 \, k\Omega$) at the bottom of the specimen and read by two Cernox thermometers. The typical temperature difference between the two thermometers was 500$\,$mK for each measurement. The temperature of the entire specimen was controlled by a film heater ($\rm 75 \, \Omega$) at the top. We cooled down this setup using liquid nitrogen and liquid helium. We measured thermal conductivity data from 3$\,$K to 25$\,$K, and at 80$\,$K. Temperatures below 4.2$\,$K were achieved by pumping out liquid helium. 

Since connectors' port of electrical cables is on the lid which makes contact with liquid helium, all electrical cables are once anchored on the top aluminum block. We used the 4-wire measurement method to measure the input power of the bottom heater. We adopted the phosphor bronze and Manganin as electrical wires for thermometers and heaters, respectively, to reduce heat extraction via wires. We estimated the amount of the heat flow via glass epoxy supports and concluded that it is small enough. Heat extraction by thermal radiation was also negligible. So we estimated that the total error in relation to heat is below 1$\,$$\%$. 

\begin{figure}[h]
\centering
 \includegraphics[width=30pc]{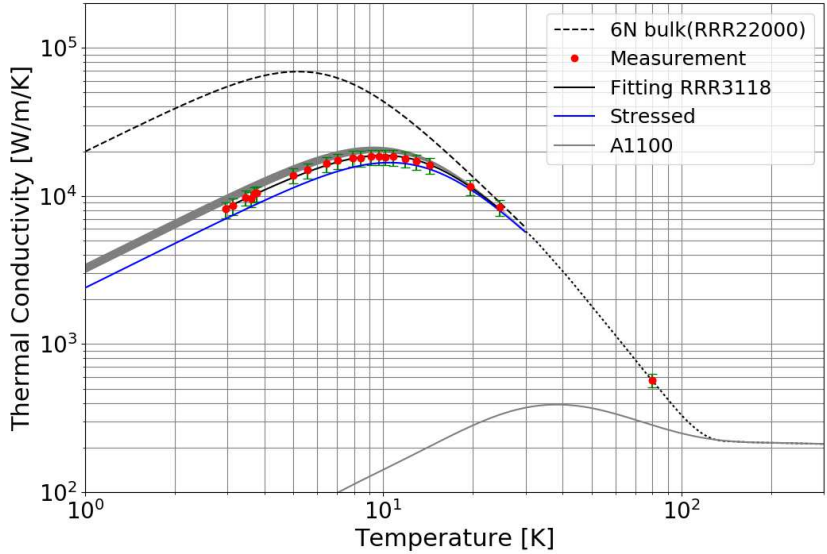}
 \caption{Red points shows the measured data, a black solid line is the fitting curve of the data by eq.($\ref{RRR_tc_estimation}$), where RRR is a fitting parameter, and a broken line is the thermal conductivity curve estimated from RRR of bulk. Grey zone represents the region of estimated thermal conductivity with 1$\,\sigma$ error of RRR results of 6$\,$N specimens. A blue line is an estimated thermal conductivity curve using eq.(\ref{RRR_tc_estimation}) for the case of deformation given by wrapping around a 10mm bar. For comparison, the thermal conductivity of industrial level pure aluminum (A1100)  is shown as a grey solid line [12].}
 \label{Thermalconductivity.png}
\end{figure}

Measured data is shown as red points in the Figure$\,$\ref{Thermalconductivity.png}. The measured maximum thermal conductivity was 18500$\,$$\rm W/m/K$ at 10$\,$K. This value is 130 times larger than the industrial level pure aluminum A1100 at 10$\,$K [12]. The black solid line shows the fitting curve of measured data by eq.(\ref{RRR_tc_estimation}), where RRR value is fitting parameter. Grey zone represents the region of thermal conductivity estimated with 1$\,\sigma$ error of RRR results of 6$\,$N specimens. Blue solid line shows estimated curve of the thermal conductivity with strain when wrapped around a 10$\,$mm bar from the RRR measurement.

In Figure$\,$\ref{Thermalconductivity.png}, we confirmed that eq.(\ref{RRR_tc_estimation}) can correctly predict thermal conductivity from the RRR. At 10$\,$K, there was only a 10$\,\%$ difference between the measured value and the estimated value derived from eq.(\ref{RRR_tc_estimation}).

\subsection{Spring Constant}
\label{Spring Constant}

Developing a heat link with a small spring constant is one of our motivations for this study because it is an important factor in reducing vibration to experimental devices. We evaluated the spring constant by measuring the resonant frequency. We compared two samples with the same cross-sectional area; stranded cable with 45 thin wires and a thick single wire with 1$\,$mm diameter.

\begin{figure}[h]
\centering
 \includegraphics[width=20pc]{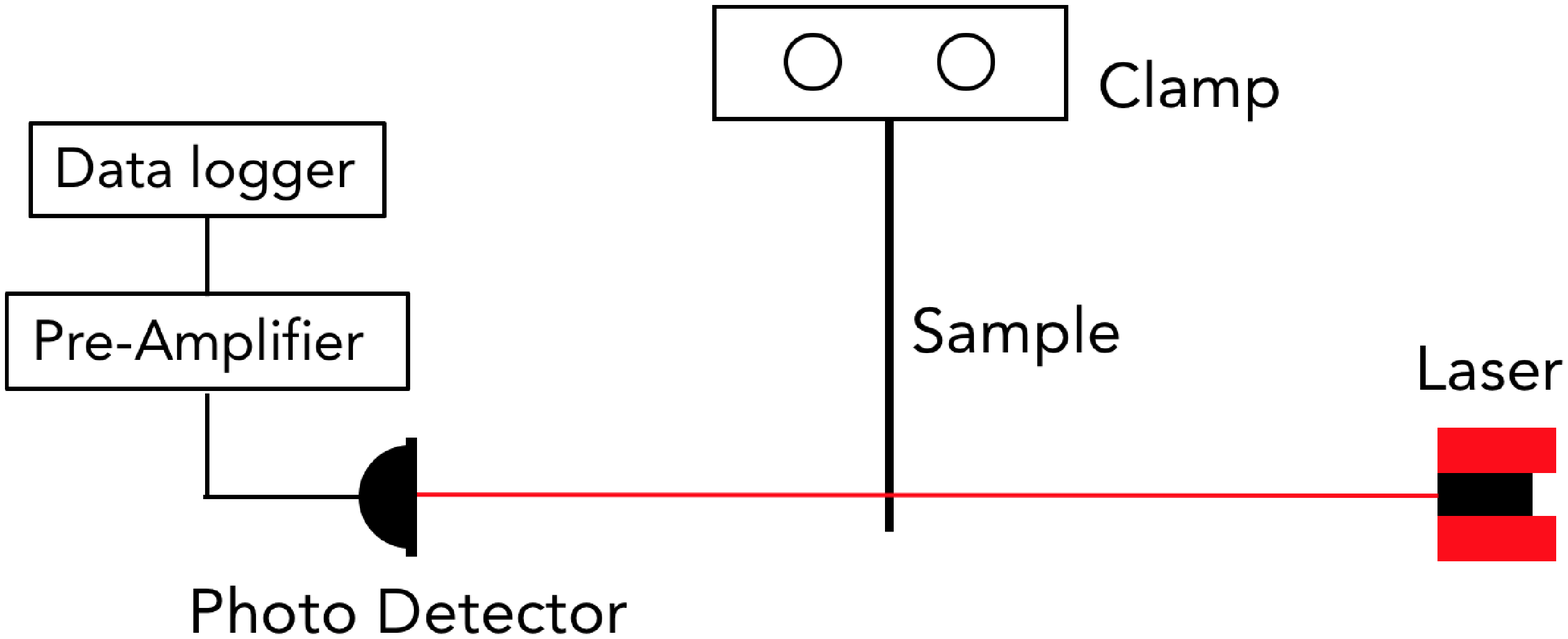}
 \caption{Shadow sensor method. Length of sample is 100$\,$mm.}
 \label{springconstant.png}
\end{figure}

We employed the shadow sensor method to measure mechanical resonant frequency as shown in Figure$\,$\ref{springconstant.png}. The sample was set on the laser beam path so that its shadow was thrown upon the photo detector. By exciting oscillation of the sample as cantilever by hand, we can measure resonant frequency of the sample from the change of detected laser power. Since high purity metal is very soft, excitation of amplitude must be small enough that it does not cause plastic deformation.

Table$\,$\ref{resonant_jikken} shows the list of measured data. We found that the resonant frequency of a strand was 1/6.5 of that of a single wire. This is equal to 1/43 decrease in the spring constant. The spring constant will be decreased 1/45 according to eq.(\ref{equation_springconstant_decrease}), therefore we conclude that the measured value is almost consistent with theory.

\begin{table}[h]
  \begin{center}
    \begin{tabular}{|c||c|c|c|} \hline
         & Cross sectional area&$\ \ \ $5\,N$\ \ \ $&$\ \ \ $6\,N$\ \ \ $  \\ \hline \hline
      Strand & $0.8 \,\rm mm^{2}$&9.6$\,$Hz & 9.8$\,$Hz \\ \cline{1-4}
      $\phi \,$1\,mm &$0.8 \,\rm mm^{2}$& 64$\,$Hz & 64$\,$Hz \\ \cline{1-4}
    \end{tabular}
    \caption{Results of resonant frequency measurement. There is 0.2$\,$ Hz difference between 5$\,$N and 6$\,$N strands. We speculate that it is because we re-strand wires by hand to decrease numbers of wires from 49 to 45. }
    \label{resonant_jikken}
  \end{center}
\end{table}

\subsection{Consideration}
\label{Consideration}

Here, we must remember to take into consideration the size effect. We cannot expect the same heat transfer from a stranded cable that consists of thin wires with a single thick wire, even if both cables have the same cross-sectional area. So we need more stranded cables to have same heat transfer with thick single wire. In order to achieve the same amount of heat transfer, the number of thin wires in the stranded cable must be increased by $\kappa / \kappa_{size}$, where $\kappa$ is the thermal conductivity without the size effect and $\kappa_{size}$ is the size effect. Then, the spring constant of N wires with the size effect, $k_{N, size}$ has a relationship with eq.(\ref{equation_springconstant_decrease}) as 

\begin{equation}
k_{N, size} = \frac{\kappa}{\kappa_{size}}\,k_{N}=\frac{1}{N}\frac{\kappa}{\kappa_{size}}\, k.
\label{spring_constant_N_size}
\end{equation}

\noindent Therefore, as long as $\frac{1}{N}\frac{\kappa}{\kappa_{size}}$ is smaller than 1, there is an advantage to using a strand cable even in cases where the size effect dominates thermal conductivity.

\section{Conclusion}
\label{Conclusion}

We developed a soft and a high thermal conductive heat link at cryogenic temperature by using high-purity aluminum (99.9999$\,$\%, 6N) thin wires. We realized small spring constant by using a stranded cable that is made of many thin wires for keeping large thermal conductance. Although size effect can limit thermal conductivity and spring constant of the stranded cable made of thin wires, we showed that there still is an advantage to use this stranded cable to reduce transfer of vibration to experimental devices. Therefore, we conclude that this approach plays an important role in devices requiring small vibration at cryogenic temperature.

\section*{Acknowledgements}
\label{Acknowledgements}

We would like to express our appreciation to Sumitomo Chemical Co., Ltd., Noge Electric Industries Co., Ltd., Arai Co. Ltd. for fabricating the heat link. We thank Mr. Hiroaki Hoshikawa and Mr. Akira Nagata in Sumitomo Chemical Co., Ltd. for useful advise in measurements. We would like to thank Prof. Akira Yamamoto and Prof. Michinaka Sugano in High Energy Accelerator Research Organization for the discussion regarding experiments.

This work was supported by MEXT, JSPS Leading-edge Research Infrastructure Program, JSPS Grant-in-Aid for Specially Promoted Research 26000005, JSPS Grant-in-Aid for Scientific Research on Innovative Areas 2905: JP17H06358, JP17H06361 and JP17H06364, JSPS Core-to-Core Program A. Advanced Research Networks, JSPS Grant-in-Aid for Scientific Research (S) 17H06133, the joint research program of the Institute for Cosmic Ray Research, University of Tokyo, National Research Foundation (NRF) and Computing Infrastructure Project of KISTI-GSDC in Korea, Academia Sinica (AS), AS Grid Center (ASGC) and the Ministry of Science and Technology (MoST) in Taiwan under grants including AS-CDA-105-M06, the LIGO project, and the Virgo project.








\end{document}